\documentclass{sig-alternate}

\usepackage{url}
\usepackage{subfigure}
\usepackage{array}
\usepackage{fancyvrb}
\fvset{frame=single,samepage=true,fontfamily=courier,numbers=left,numbersep=2pt,baselinestretch=0.50,fontsize=\small}

\begin{document}

\title{Improving Accessibility of Archived Raster Dictionaries of Complex Script Languages}

\numberofauthors{3}

\author{
\alignauthor
Sawood Alam\\
\affaddr{Dept. of Computer Science}\\
\affaddr{Old Dominion University}\\
\affaddr{Norfolk, Virginia, USA}\\
\email{salam@cs.odu.edu}
\alignauthor
{Fateh ud din} B Mehmood\\
\affaddr{National University of Sciences and Technology}\\
\affaddr{Islamabad, Pakistan}\\
\email{fatehh@gmail.com}
\alignauthor
Michael L Nelson\\
\affaddr{Dept. of Computer Science}\\
\affaddr{Old Dominion University}\\
\affaddr{Norfolk, Virginia, USA}\\
\email{mln@cs.odu.edu}
}

\maketitle

\begin{abstract}

We propose an approach to index raster images of dictionary pages which in turn would require very little manual effort to enable direct access to the appropriate pages of the dictionary for lookup. Accessibility is further improved by feedback and crowdsourcing that enables highlighting of the specific location on the page where the lookup word is found, annotation, digitization, and fielded searching. This approach is equally applicable on simple scripts as well as complex writing systems. Using our proposed approach, we have built a Web application called ``Dictionary Explorer'' which supports word indexes in various languages and every language can have multiple dictionaries associated with it. Word lookup gives direct access to appropriate pages of all the dictionaries of that language simultaneously. The application has exploration features like searching, pagination, and navigating the word index through a tree-like interface. The application also supports feedback, annotation, and digitization features. Apart from the scanned images, ``Dictionary Explorer'' aggregates results from various sources and user contributions in Unicode. We have evaluated the time required for indexing dictionaries of different sizes and complexities in the Urdu language and examined various trade-offs in our implementation. Using our approach, a single person can make a dictionary of 1,000 pages searchable in less than an hour.

\end{abstract}




\section{Introduction}

\begin{figure*}[!t]
\begin{center}
\subfigure[BookReader: Search for an Urdu term returned no matches due to lack of OCR support although the term is present in the book.]{\label{img:brsearchur}\includegraphics[width=17.6cm]{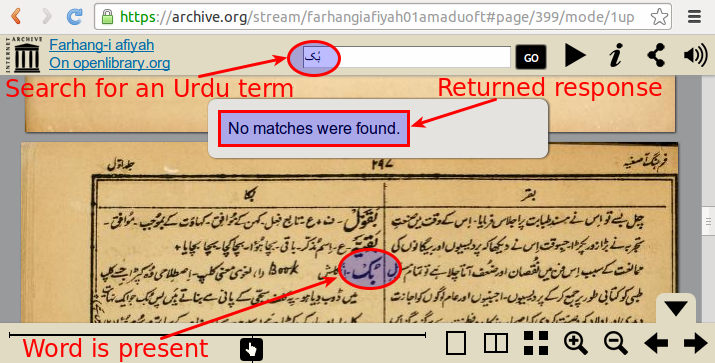}}
\subfigure[BookReader: Search for a common English term returned too many matches, making it difficult to find the term definition in the dictionary.]{\label{img:brsearchen}\includegraphics[width=17.6cm]{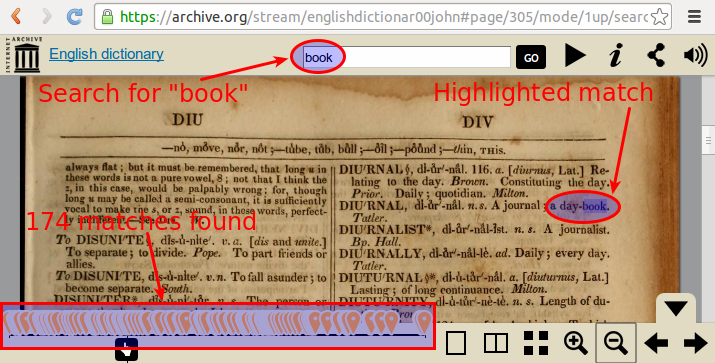}}
\caption{BookReader: Problems in searching words in raster dictionaries.}
\label{img:brsearch}
\end{center}
\end{figure*}

The Internet Archive (IA) and Open Library offer over 6 million fully accessible public domain eBooks~\cite{archtext}. Among these eBooks there are handful of Unicode eBooks (for example, the Project Gutenberg collection~\cite{gutenberg}), but most of them are PDFs and scanned raster image based eBooks. To make these books accessible from a Web browser, IA uses an open-source application called BookReader~\cite{bookreader}. To make the content of these raster eBooks searchable, IA uses ABBY FineReader~\cite{abbyocr} for Optical Character Recognition (OCR)~\cite{dhisttext}. For full-text searching, OCR engine returns page numbers and coordinates of the surrounding box, that is then used by the BookReader to annotate and highlight appropriate regions on corresponding pages. ABBY FineReader claims support for up to 190 languages~\cite{abbylang}. Unfortunately several major languages such as Indian languages (Hindi, Punjabi, Telgu, Marathi, etc.) and complex script languages (Urdu, Persian, Sindhi, etc.) are not supported by this OCR engine. As a consequence, scanned books in these languages do not support full-text searching as illustrated in Figure~\ref{img:brsearchur}.

Among these archived books there are a handful of dictionaries in various languages, ranging from very rare and classical to modern dictionaries. These dictionaries accumulate a treasure of ancient and obsolete as well as modern and contemporary words and phrases that are of equal interest to archivists and linguistics. In dictionaries, fielded searching is more important and desirable than full-text content searching. Generally, lookup in a dictionary involves a word or phrase, part of speech, origin language and other related metadata. Searching in the definition field is rarely desired. The type of OCR-powered full-text searching currently available in IA's BookReader makes it very difficult to lookup definitions of common words in a dictionary. For example, if we search for the term ``book'' in the scanned copy of~\cite{samueldict}, it returns 174 matches on various pages including pages where the term has appeared in the definitions or examples as illustrated in Figure~\ref{img:brsearchen}.

To enable easy lookup, traditional dictionaries have a distinct property of being a sorted index of words that makes the interaction with them different from other books. One can easily use binary search to flip through pages while looking at the header of the pages to locate the desired page. A destination page in a dictionary may or may not have the desired word, but it ensures that if the lookup word was not found on that page then it does not exist in that dictionary. Using this property of dictionaries, we are progressively indexing pages of dictionaries manually to provide various levels of accessibility and searchability. A sparse index can make lookup in a normal size dictionary possible in matter of few hours while a complete indexing requires a few days of crowd-sourced work. Complete digitization and annotation is a long-term ongoing crowd-sourced effort.

\section{Background}

Digitizing scanned book archives is an important step towards improving accessibility. Digitization can be done manually by crowdsourcing which yields good quality, but is slow process in contrast with an OCR assisted approach which is fast and scalable but results in poor quality recognition. In complex script languages, the accuracy of the OCR is often unacceptable and sometimes there is no support at all. To understand the current state of the character recognition in Asian and complex script languages, various surveys have been published~\cite{scrrec, charrec, indicscr, arabicrec}. Arabic, Persian, Urdu, and some other languages share the same script with slight variation in character-sets and they are all written right-to-left. Persian and Urdu are very similar in terms of typography as they both are traditionally written in Nastaleeq (a complex cursive style) while Arabic is mostly written in Naskh (a complex but non-cursive style). Character segmentation and recognition is challenging in these scripts, hence there were OCR efforts that work with or without segmentation~\cite{arprocr, urocr, nastocr, arurocr, hmmtranshand}.

Urdu is a widely spoken language worldwide with a total number of 104 million speakers~\cite{urduspeaker}, with the majority in India and Pakistan. UrduWeb Digital Library (UWDL)~\cite{urdulib} is an initiative of UrduWeb~\cite{urduweb} where a designated library team manually digitizes copyright-free Urdu books and publishes them on the Web in Unicode format. UWDL has started a sub-project of digitizing dictionaries because Web-based modern dictionaries are missing historical vocabulary that are not in use. Scanned copies of various classical Urdu dictionaries are preserved in IA without any OCR processing. Digitizing these dictionaries properly with all the attributes stored separately to enable fielded searching is a time intensive task. We simplified this task by indexing images and serving appropriate scanned pages on lookup. This approach has proven successful and efficient and it is not limited to just the Urdu language.

\section{Related Work}

There have been a lot of work in preserving books in the form of scanned images, digitizing scanned pages using OCR engines and crowdsourcing, analyzing digital data, and building archive explorers~\cite{lexdhisd, dlip, cculibm, blldict, millatin, transscan, aplarchexp, ppeudl, retrogreen}. While OCR engines offer a scalable mechanism to digitize scanned images, they have limited accuracy or no support for many of the world's popular languages, hence automating digitization work-flow~\cite{dworkflow} is not feasible. Our focus is dictionaries, a very specific class of books, that are different from other general book in many ways including fielded information and some sort of ordering for easy lookup.

\subsection{Major Digital Libraries}

The Google Books initiative~\cite{gb} is promoting the democratization of knowledge by scanning the World's books and putting them online. In 2010 Google estimated the total number of books in the world about 130 million~\cite{gbcount} and intended to scan them all in a decade. Google Books has scanned over 30 million books by April 2013 but the process was since slowed down due to copyright issues~\cite{ndpl}.% Open Library and HathiTrust also mirror books from Google Books collection.

The Open Library~\cite{olib} is an initiative of the Internet Archive to facilitate access to digital books. It is a book database and digital lending library. It serves over six million digital books in various formats, but several books in the collection are in the form of scanned raster images of printed book pages also available as PDFs. Apart from downloadable books it offers online access to several scanned books with the help of a Web browser based application.

HathiTrust~\cite{hathitrust} is a collaborative repository of digital content from various research institutions and libraries to preserve cultural records and to ensure accessibility and availability of the preserved content in future.

These three libraries are good sources of scanned images of dictionaries. Mirdeghan conducted a survey to compare the interfaces of these three libraries, which shows that the majority of participants preferred Open Library followed by Google Books~\cite{libuicomp}. Our interface idea is inspired by the Open Library BookReader interface with extra features specific to dictionary exploration.

\subsection{BookReader}

BookReader~\cite{bookreader} is an open-source community collaborated Web application for online reading of books which powers the Open Library as well as several other digital libraries. BookReader supports various layouts like single page, two pages, or thumbnails of multiple pages on the viewport, zooming functionality, flipping pages forward and backward, seeking direct access to a page, auto-play to flip pages automatically, highlighting matched text on the page, read aloud, and full-text searching. To facilitate full-text searching, text-to-speech, and highlighting it uses an OCR engine to process scanned pages. Figure~\ref{img:searchapires} illustrates a response from the Open Library's search inside book API (Application Programming Interface)~\cite{searchapi}. As illustrated in Figure~\ref{img:brsearchen}, this response is used to identify matching pages, highlight matching text on those pages, and populate the tooltip in the BookReader. It is made for exploring general purpose books, hence it does not contain specific controls and features that a dictionary explorer may require.

\begin{figure}[!h]
\centering
\begin{Verbatim}
{
  "ia": "designevaluation25clin",
  "q": "\"library\"",
  "page_count": 224,
  "body_length": 475677,
  "leaf0_missing": true,
  "matches": [{
    "text": "The... {{{Library}}}... Writing...",
    "par": [{
      "page": 14,
      "page_width": 2134, "page_height": 3328,
      "b": 1090, "t": 700, "r": 2024, "l": 192,
      "boxes": [{
        "r": 1560, "b": 957, "t": 899, "l": 1378
      }]
    }]
  }]
}
\end{Verbatim}
\caption{BookReader: Search API Response}
\label{img:searchapires}
\end{figure}

\subsection{ABBY FineReader}

ABBY FineReader is a commercial OCR engine. IA uses it to process scanned images of books to power BookReader. It recognizes up to 190 languages that include natural, artificial, and  formal languages, but unfortunately out of the world's 100 most popular languages (based on number of native speakers)~\cite{langspeak} 63 are not supported. This proportion of unsupported languages roughly holds true for top 50 (29 unsupported) and top 10 (4 unsupported) languages as well. Also the distribution of the accuracy of recognition over supported languages is not mentioned in~\cite{abbylang}. Figure~\ref{img:brsearchur} shows an unsuccessful search for an Urdu term in a dictionary using the Open Library BookReader due to the lack of support of OCR in the Urdu language, although the term exists on the page. In the case of a dictionary, users primarily want to search in the words not in the definitions, but this OCR engine cannot distinguish definition text from the lookup word.

\subsection{Digital Dictionaries of South Asia}

University of Chicago has worked on a project to digitize dictionaries of South Asian languages~\cite{ddsa}. They have selected 26 modern literary languages of South Asia and selected at least one multilingual dictionary for each language. They have also chosen a monolingual dictionary for more frequently taught languages. They have digitized a handful of dictionaries manually and created an interface for lookup and perform fielded searching in them. It requires a lot of focussed manual labour to digitize books this way and the process is not scalable. Most of their work is available under a Creative Commons license~\cite{cc} but some materials have further restrictions.

\subsection{Open Annotation}

Open Annotation Data Model~\cite{openannotation, annotpersist, annothist} allows users to associate additional information with an existing Web resource or a specific part of it. It is called open because the model utilizes Linked Data~\cite{ld} principles to connect open Web resources together. This technique can be used to annotate words on scanned images of dictionary pages, interlink two words across dictionaries, associate user comments about the word, or store digitized representation of the page segment related to the word.

\subsection{Distributed Proofreaders}

Distributed Proofreaders~\cite{dproof, dp} was developed in 2000 to digitize Public Domain books which is primarily used by the Project Gutenberg~\cite{pgb}. It is a crowdsourcing utility to allow users to collaborate on digitizing scanned books one page at a time. Users are presented with the original scanned image of the page and an editable text which is pre-populated with the OCR engine generated text if possible, which users can then edit to correct mistakes. This concept can be used in the process of annotation and digitization of dictionaries.

\subsection{Urdu Encyclopedia}

Urdu Encyclopedia~\cite{uenc} is an UrduWeb knowledge-base project with currently over 330,000 pages primarily containing digitized dictionaries of general terms in the Urdu language with some additional specialized subjects. This encyclopedia was created by accumulating licences and digital data from various publishers. We are using this as one of the sources of data in our implementation.

\section{Methodology}

We have adopted a progressive approach which starts with little effort and improves the accessibility as more energy is put into the system.

\subsection{Indexing}

\begin{figure}[ht]
\centering
\includegraphics[width=\linewidth]{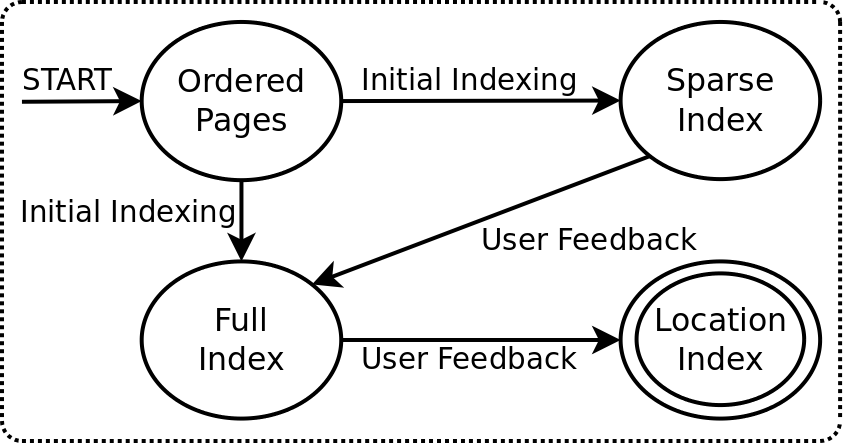}
\caption{Index State Transition}
\label{img:idxstate}
\end{figure}

To enable lookup in a dictionary, an index of words needs to be prepared that points to the corresponding page numbers in the dictionary. We have chosen a progressive approach for indexing, so that basic lookup feature can be enabled very rapidly and further improvements can be made later to increase the accessibility for the raster dictionary. Figure~\ref{img:idxstate} illustrates the transition among various states of the indexing process that a typical dictionary may go through. Each indexing state is described in detail below.

\subsubsection{Ordered Pages}

Indexing starts with the scanned pages organized in the order they appeared in the printed book. Dictionaries in IA's BookReader are present in this ``ordered pages'' state. This state allows lookup of a word similar to how we interact with printed dictionaries. The user opens a page with a guess then compares the header of the page with the lookup word and decides, should it be flipped few pages forward or backward and repeats this process until reaches the destination page. On the destination page either lookup word is present or it does not exist in the entire dictionary.

Scanned books have a common issue of missing pages. If the book is rare then not much can be done about it. One strategy that we use in such cases is to add a custom page to let the users know that the page is missing, so that they can contribute if they have a copy of the book. For the sake of easy binding and portability, large books are often published in multiple volumes. In general purpose books this is not an issue, but in case of a dictionary, it is desired to have all the volumes together. In our testbed, a dictionary of four volumes was combined to form a single book. We found that those four volumes in the archive were not from the same publication and they had misaligned volume partition. As a consequence, the combined dictionary contained several duplicate and missing pages (which were later added).

\subsubsection{Sparse Index}

Sparse indexing is the quickest way to improve the lookup experience. With minimal effort, manual lookup can be turned into automated lookup process. To prepare a sparse index, a list is created that contains the first words of every page along with the corresponding page number. This approach is only applicable if the dictionary words are already sorted, hence the list is sorted by words as well as page numbers. While this is true in most cases, section~\ref{nesting} describes some special cases where all the words in a dictionary are not sorted as a flat list of words. The process of lookup involves finding the appropriate place in the list where the lookup word can be inserted without breaking the sorting. In other words, finding a word in the sorted list which is either same as or prior to the lookup word, but the next word (if there is any) in the list is after the lookup word when performed string sorting. Page numbers associated with the word matching this criteria indicate where there is the possibility of existence of the lookup word. A sparse index cannot ensure the presence or absence of the lookup word in a dictionary, but it ensures that if the word exists in the dictionary, it has to be on the returned page.

A sparse index can also be made by listing only the last words of each page instead of the first words. In that case, only the matching criteria changes a little, but overall approach remains the same. This variation does not add any value, except it might be a more expressive and readable way to write the lookup algorithm and it might also be helpful in fine-tuning an implementation for performance reasons.

Not all the pages in a dictionary have one or more words, sometimes the definition of a word and related examples span over multiple pages. This does not invalidate the working of this mechanism but at the time of implementation, this knowledge can be helpful. For example if one decides to use a plain array for storing sorted list of sparse index words and reuse array index to determine the corresponding page number implicitly, then knowing this fact is essential in order for the implementation to work properly.

A sparse index does not necessarily require indexing each page. One may decide to select a window of $N$ pages and only add one entry per $N$ pages in the index. This reduces the time required to index the whole dictionary, but shifts the burden to the user to scan for the word in the given window of pages. Additionally, it is not necessary to choose a window size of fixed number of pages. A practical example could be to index only the pages where a new letter of the alphabet begins. It only requires index of size equal to (or less than) the size of the alphabet in the given language. There are variable number of words starting from each letter and they span over varying number of pages. Such an index can only help the user to jump to the start of the section where first letter of the lookup term begins.

A sparse index is a sorted list hence almost any searching algorithm optimized for sorted list can be used for lookup in the list. In practice these sparse indexes are small lists, hence a linear searching algorithm may prove to be as good as a binary search.

\subsubsection{Full Index}

Full indexing involves preparing a list of all the words in the dictionary along with their corresponding page numbers. Unlike a sparse index, it does not require the list to be in any explicit sorted order. It even works on dictionaries which do not arrange their words as a flat sorted list. There are some dictionaries that arrange words based on their roots and all the variations from that root are internally sorted but there is no flat list ordering. In some dictionaries, words made of common prefixes or suffixes are accumulated together and break the overall sorting. In such cases, a sparse index is not an option. A full index has an added advantage over a sparse index in the way that it ensures either presence or absence of the lookup word in the dictionary.

Creating a full index manually takes longer as compared to a sparse index. In one of the classic dictionaries we worked on search words were not always in the beginning of the line, instead they were all over the page more like a paragraph, mixed with the definitions. To distinguish words from meaning, the dictionary used underline markers. Such situations may make the process of manually indexing dictionaries difficult and error-prone.

We propose an alternate crowd-sourced slow approach of full indexing, especially for large dictionaries. In this process initially a sparse index in made. As users search for words in the dictionary, they are asked for the feedback if they could find what they were looking for on this page. Their feedback is recorded. Given enough time, a full index will eventually be prepared. To avoid malicious feedback or unintentional error, a threshold of minimum number of agreements can be set. Alternatively, a democracy model can be used that is utilized in various crowd-sourced software localization systems, in which a translation with maximum votes wins.

Full indexes may or may not be in any sorted order as this is not a requirement for them to work. Hence a linear search is the easiest to implement. But if the list of words is large then the index can be sorted explicitly to enable binary search. Sorting or rearranging a full index does not affect the lookup functionality hence more advanced data structures like tries~\cite{trie} can be built for efficient lookup.

\subsubsection{Location Index}

A sparse or full index can lead the user to the appropriate page, but the user has to locate the lookup word manually on the page. We can improve the accessibility here as well and point user's attention to the exact coordinates of the image where the lookup word exists. A small highlighted sticker, pin, or marker can be used as an overlay on the image to precisely locate the word on the page.

Location indexing is best suited as a crowd-sourced process in which users looking for a word in a page are asked to help position a marker on the page (if the word exists on that page). Users are allowed to move the existing markers around to correct the location. If multiple attempts are made to correct the location of a word, for simplicity, the last attempt can win. To avoid spamming a linear or quadratic mean can be utilized to determine the coordinates of the location marker. This is a finite task and given enough time, placement of markers will eventually be completed.

\subsection{Annotation}

Indexing is good for easily finding information that is in the scanned dictionaries. But we can improve the usefulness of a dictionary by attaching additional related information, resources, and comments in the form of annotations. Interacting with the location marker of the word reveals attached annotations in an overlay panel.

Annotation is a crowd-sourced process in which users looking for a word in a page are encouraged to annotate the word. Users are allowed to post comments and related resources in the overlay panel associated with the location marker. Unlike indexing, annotation is an ongoing process that will never reach an end, but it will mature as time passes and more people interact with the system.

\subsection{Digitization}

Digitization is the final stage of improving the accessibility and usefulness of a scanned raster dictionary. A dictionary in this stage is able to give users access to a fielded representation of every lookup word along with the definition, part of speech, examples, and other metadata in Unicode.

In the process of digitization, users fill a fielded form with the word and all its attributes as present in the image (if the word is not correctly digitized already). Related regions on the scanned pages are highlighted using rectangles. Often there is more than one rectangle to cover all the related text which might be in different columns on the same page or continue in the next page. This related region highlighting can be automated if the location indexing is completed. An aggregated analysis of the location marker positions over all the pages of a dictionary helps determining margins and number of columns on each page. Combining this information with the location of two consecutive words gives the coordinates of all the regions related to the first word. If the location indexing is not complete or the automatic location of related region estimates are not accurate then users are encouraged to correct it manually while digitizing.

Curation techniques can be used to check the correctness of the digitization. This information can be stored with the reference of page number and dictionary identifier. Given enough time dictionaries will be fully digitized in a way that will facilitate fielded searching. As a by-product, combining digitized data with coordinates of rectangular boxes can also be used for training OCR engines.

\subsection{Sorting}

Sparse indexing relies heavily on the fact that the dictionary words are arranged in a sorted order in the scanned images or PDF. Unfortunately, not every dictionary in every language follows linear sorting. Some dictionaries use different mechanisms of lookup in which the lookup key is not necessarily a prefixed substring of the desired word. Derivational prefixes and suffixes also affect the ordering.

\subsubsection{Unicode Collation}

Collation is ordered assembly of written information under a standard ordering scheme for easily finding an item in a list of items. In many languages, Unicode character values are not in the alphabetical order. For example Arabic script is used for many other Asian and African languages such as Arabic, Persian, Urdu, and Pashto~\cite{arabicscript}. These languages inherit the basic alphabet set from Arabic then add or remove letters. In the Unicode table Arabic script reserves values from \texttt{0600} to \texttt{06FF}. In this table, characters from the Arabic language alphabet are mostly sorted with a few exceptions. All other languages that use Arabic script as their base have their additional characters below these basic characters. This behavior changes the ordering of characters in the table from their alphabetical order. To solve this problem, the Unicode collation algorithm~\cite{uca} was introduced. Unicode has also introduced a project called Common Locale Data Repository~\cite{cldr} to bring full support of locale related issues in the world's languages.

Sometimes dictionaries of the same language do not agree on a single ordering scheme. To deal with this issue, custom sorting functions can be written for every dictionary or a set of dictionaries that agree on the same ordering scheme.

\subsubsection{Compound Letters}

Some languages have compound letters where two or more characters are combined to form a single phoneme. Dictionaries often disagree on the matter of compound characters. For ordering purpose some dictionaries treat the compound character as a single letter, while others treat them as combination of multiple letters. For example in the Urdu language there is a character with Unicode value \texttt{06BE (ARABIC LETTER HEH DOACHASHMEE)} which has a place in the alphabet but it is not used independently. It can be used as a modifier to a dozen other letters to change the sound of the phoneme. Some Urdu dictionaries consider those compound letters as independent letters while others consider them as two separate characters while sorting the list of words. Accented characters and diacritic marks also affect the sorting order. A custom sorting function needs to consider these aspects to work properly.

\subsubsection{Nested Ordering}
\label{nesting}

In some dictionaries, lookup is not provided by words but a high-level search key is used instead. For example in some Arabic language dictionaries, root words are chosen as a key. Top level sorting is done on root words. All the derivative forms of the root words are sorted alphabetically on a secondary level. This approach keeps related words close and form a cluster that share similar meaning. This type of organization helps users understand the meaning of words in better context. This benefit comes with a cost as it requires the user to know the root of the lookup word in advance. For example a root word in Arabic of the form \texttt{L1-L2-L3} can have derivatives in the form of \texttt{E*-L1-E*-L2-E*-L3-E*}, where L1, L2, and L3 are three letters of the root word while E* is zero or more occurrences of extra letters in the word. Sometimes derived words change or remove root letters to simplify pronunciation (under some grammar rules) that makes identifying root of a derived word complex.

Another example of nested sorting and clustering words is found in some dictionaries where words that have same derivational prefix or suffix are grouped together. These compound words are sometimes written as a single word, but sometimes have spaces (or other word-boundary characters like a dash in other languages). Often there is no consensus about one way of writing or the other.

\subsection{Retrieval}

\begin{figure}[!t]
\centering
\begin{Verbatim}
{
  "query": "SEARCH TERM",
  "language": "LANGUAGE CODE",
  "resources": [{
    "type": "RESOURCE TYPE",
    "href": "RESOURCE URL",
    "meta": {
      "contributor": "NAME",
      "updated": "DATE",
      "OTHERFIELDS": "THEIR VALUES"
    }
  }],
  "definitions": [{
    "text": "DEFINITION TEXT",
    "meta": {
      "contributor": "NAME",
      "updated": "DATE",
      "OTHERFIELDS": "THEIR VALUES"
    }
  }],
  "dictionaries": [{
    "id": "DICTIONARY ID",
    "exists": "yes/no/maybe",
    "pages": [{
      "number": "PAGE NUMBER",
      "src": "URL OF THE PAGE IMAGE",
      "width": "WIDTH OF THE IMAGE",
      "height": "HEIGHT OF THE IMAGE",
      "location": {
        "x": "X-COORDINATE",
        "y": "Y-COORDINATE"
      },
      "boxes": [{
        "top": "TOP Y-COORDINATE",
        "bottom": "BOTTOM Y-COORDINATE",
        "left": "LEFT X-COORDINATE",
        "right": "RIGHT X-COORDINATE"
      }],
      "annotations": [{
        "id": "ANNOTATION ID",
        "text": "ANNOTATION TEXT",
        "meta": {
          "contributor": "NAME",
          "updated": "DATE",
          "OTHERFIELDS": "THEIR VALUES"
        }
      }]
    }]
  }]
}
\end{Verbatim}
\caption{Search API Response}
\label{img:farhangapires}
\end{figure}

Figure~\ref{img:farhangapires} illustrates a response that enables the client to represent the definition of the lookup term from various sources, related resources, different dictionary images with coordinates to locate the word on the page, locations of the bounding boxes that contain the definition and other related information, and annotations for the word. Lines 4--12 have list of related resources such as descriptive images or audio. Lines 13--20 hold a list of definitions and examples of the word from external Web services or user contributions in Unicode. Lines 21--49 have data specific to scanned dictionaries. If the lookup language has multiple associated dictionaries then this block of response object will have an array of all the associated dictionaries, each dictionary internally contains related pages, and each page has source, dimensions, coordinates of the word location, coordinates of the bounding boxes, and user contributed annotations. Array blocks in the response object may have zero or more elements depending on the available data. Line 23 can have ``yes'' or ``no'' as its value for fully indexed dictionaries and in case of sparsely indexed dictionaries, the value can be ``yes'' or ``maybe''. It is helpful to return a relevant page even if the word is not found on the page, this gives the user confidence and an opportunity to provide feedback in case of an error.

\section{Reference Implementation}

\begin{figure*}[!th]
\begin{center}
\subfigure[Searching an English dictionary with location index.]{\label{img:deen}\includegraphics[width=17.6cm]{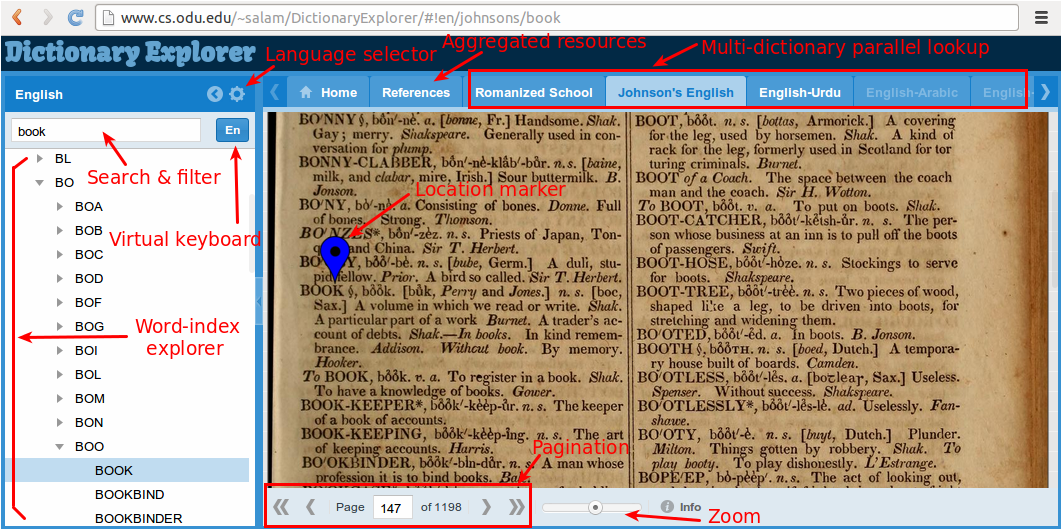}}
\subfigure[Searching an Urdu dictionary with location index and annotations.]{\label{img:deur}\includegraphics[width=17.6cm]{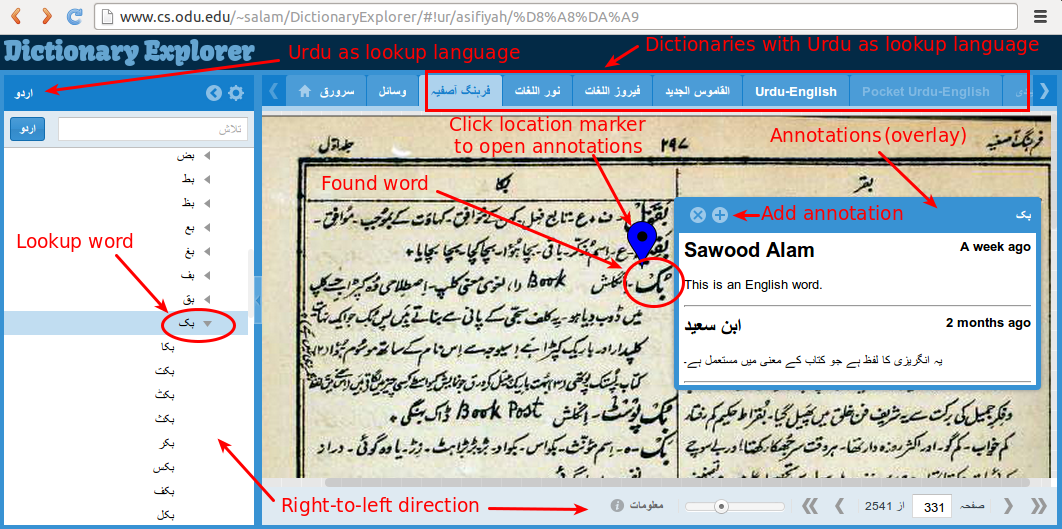}}
\caption{Dictionary Explorer: Multilingual and multi-dictionary lookup with highlighted features.}
\label{img:de}
\end{center}
\end{figure*}

Using the techniques described above, we have created an application called ``Dictionary Explorer'' (Figure~\ref{img:de}) for UrduWeb Dictionary project. We have added a handful of monolingual and multilingual dictionaries in various languages that are in different indexing states. Dictionary Explorer is built using open-source Web technologies like PostgreSQL~\cite{psqlintro, pracpsql} as the database engine, Ruby on Rails~\cite{agileror} as Athe PI endpoint, and ExtJS~\cite{extjsaction} for its client side user interface. Our implementation respects various needs of multilingual and bidirectional interfaces~\cite{bilingpref, hlpucorp}.

\subsection{Multilingual Multi-dictionary Lookup}

Dictionary Explorer offers a way to select one of many available lookup languages and provides indexes based on the selected language. Every language is associated with a subset of all the available dictionaries. Lookup for a word in any language enables tabbed interface for associated dictionaries and loads appropriate pages in each dictionary simultaneously. If that word is annotated in some dictionary pages, those annotations appear on appropriate pages. Dictionary Explorer has a dedicated tab for accumulating third party Unicode online dictionary results, related resources, audio, and user contributed materials.

\subsection{Searching and Exploring}

Dictionary Explorer provides various means to find words in various dictionaries including traditional manual page flipping, exploring with the help of a tree-style word index, and searching by entering the lookup word in a search box. The tree-style word index is dictionary independent presentation that is built using large corpora of words in each of the available lookup languages. It helps finding a word in dictionaries without the need of appropriate input methods or keyboards. The search box is accompanied by a language-sensitive on-screen-keyboard to help users type the lookup words. Typing the lookup term in the search-box filters the word index in real-time.

In our experiments, we found that the depth of three letters is practically suitable for the tree to filter the list down to a manageable number of words. After that we load the actual words as leaf nodes without further nesting. This optimal depth might be different for some languages, but this is typically used by various printed dictionaries of many languages as they write three letter index on the header of each page. The explorer also comes with a full-featured paging toolbar and zooming functionalities.

\subsection{User Contribution and Feedback}

Context-sensitive user interface elements, appropriate notifications, and intuitive interface of Dictionary Explorer encourages users to contribute by providing feedback about presence or absence of the word on the page, locate the word on the page~\cite{locatetext}, annotate the word by adding comments and linking with other resources, and digitize the dictionary. This ongoing collaborative effort improves the accessibility of the scanned dictionaries.

\section{Evaluation}

Our progressive approach of indexing involves some degree of human input at each level. A language that has a very little or no online community may not get any added benefit from this. For such dying languages~\cite{dlminority, orallit}, Dictionary Explorer is still at least as good as IA's BookExplorer.

\subsection{Indexing Time}

We have started our evaluation with a small (180 pages) English-to-Urdu (Romanized) dictionary~\cite{romandict}. The lookup words were in English, hence associated OCR processed text format of the dictionary was also available. Using regular-expression we cleaned the text except the first lookup word appearing on each page then manually corrected some errors. This process took us less than ten minutes and a sparse index for the dictionary was created.

In the next phase we have chosen an old monolingual Urdu dictionary~\cite{asifiyadict} that has four volumes with over 2,500 pages combined. Since the process of creating a sparse index requires frequent page flipping, two people have decided to work together, one person was flipping pages and pronouncing the first word of each page while the other was typing it. This approach has given us a rate of indexing of over 25 pages per minute. To simplify the process of indexing, later we created a small Web-based utility that provides a text-area for typing words (one entry per line) and scanned page appearing next to it. This utility keeps track of lines and every time enter key is hit or the cursor moves up or down, it loads the appropriate scanned page. Using this utility, we were able index over 20 pages per minute per person.

Finally, we chose another classical monolingual Urdu dictionary~\cite{noordict} that has four volumes with over 3,200 pages combined. Due to the clustering of derived words, this does not have an overall linear ordering. In such a case a sparse index would not be very helpful, hence we decided to build a full index as the initial indexing process. We distributed small subsets of pages among UWDL members and asked them to index all the lookup words appearing in each page in their free time. This process was completed in 60 days with the help of 13 volunteers, who have indexed over 75,000 words and phrases combined. We have measured an indexing speed of about 8--10 words per minute per person. This mileage may vary based on complexity of the script, available input methods, readability of scanned documents, typing speed of contributors and various other factors. Distribution and tracking of pages for indexing was handled manually in this case, but utilities like Distributed Proofreaders can be used to automate the process of crowdsourcing.

\subsection{Index Placement}

A sparse index is usually small in size (say, a dictionary with 500 pages will have a flat list of 500 words in its index) and can be transported to the client side easily, but a full index is several fold larger than a sparse index because it depends on the number of words in the dictionary, not the number of pages. The benefit of a client side index is that it requires very little communication from the server during subsequent lookup of words. In our Dictionary Explorer application, we had multiple languages and every language was associated with multiple dictionaries. If we load the index on the client side, there will be several indexes (one for each language) and the combined size will make the initialization of the application slow, hence we keep the index on the server and return a combined response (Figure~\ref{img:farhangapires}) for each lookup that has pointers to all the dictionaries associated with the lookup language.

\subsection{Prefix Index}

For the ease of lookup, we have added a tree-like explorer in our application which is language specific but dictionary independent. It is built using three letter unique prefixes of all the words from a spell checker word list for the language. Using prefix reduces the data required to build the tree-like explorer to an acceptable size and splits the entire word list in manageable chunks. When a third letter is expanded in the tree, it loads a list of words from the server with that three letter prefix and appends them as leaf nodes under that third letter. For performance reasons, we cache the list on client side so that subsequent lookups under the same prefix do not load the list from the server again.

\begin{table}[!t]\scriptsize
  \centering
  \caption{Prefix analysis of 144,106 English words.}
  \label{tab:prefix}
  \begin{tabular}{>{\raggedleft}p{0.4cm}|>{\raggedleft}p{0.8cm}>{\raggedleft}p{0.4cm}>{\raggedleft}p{0.8cm}>{\raggedleft}p{0.6cm}>{\raggedleft}p{0.9cm}>{\raggedleft}p{0.85cm}>{\raggedleft\arraybackslash}p{0.75cm}}
    \hline
    \textbf{Size} & \textbf{Count} & \textbf{Min} & \textbf{1st Q} & \textbf{Med} & \textbf{Mean } & \textbf{3rd Q} & \textbf{Max}\\
    \hline
1 & 26 & 121 & 2,753 & 5,010 & 5,503.0 & 7,603.0 & 15,620\\
2 & 528 & 1 & 2 & 11 & 271.0 & 270.5 & 5,062\\
3 & 3,995 & 1 & 1 & 7 & 35.8 & 33.0 & 1,798\\
4 & 18,026 & 1 & 1 & 2 & 7.9 & 7.0 & 753\\
5 & 40,927 & 1 & 1 & 1 & 3.5 & 3.0 & 616\\
6 & 62,767 & 1 & 1 & 1 & 2.3 & 2.0 & 227\\
    \hline
  \end{tabular}
\end{table}

Table~\ref{tab:prefix} shows standard statistical distribution of words and how the size of index changes with the size of prefix. This table was generated using 144,106 English words from the Ispell word list~\cite{ispell}. We found similar statistical distributions in other languages as well. If only the first letter of each word is indexed, there will be only 26 index entries, but each index will point to an average of 5,000 words and the maximum number will go over 15,000. A three letter prefix is optimal, because generating prefix indexes of size four or more increases the number of entries in the index to a limit which is not suitable for transferring to the client side. There are a few outliers that increase the maximum number of words that a three letter prefix can have. These are mainly derivational prefixes like ``con'', ``dis'', ``pro'', ``pre'', and ``int''. A more intelligent approach of generating balanced prefix buckets is to have a tolerance limit of maximum number of words a prefix can have, then split larger buckets by increasing the prefix size for those set of words only.

Currently this prefix index only helps initiating lookup, but it can have pointers to pages of various dictionaries of that language. This will help minimise search requests while expanding the tree, but the data structure for the tree will become complex and large. It will also be coupled with the dictionaries, which will force regeneration of the data structure every time a new dictionary is added. An alternate approach to generate such an explorer is to generate all possible next level combinations based on letters of the alphabet of the language when a node is expanded. This approach requires no knowledge of any word list except the letters of the alphabet, hence it is very lightweight in size. But this approach yields many branches that lead to no meaningful words.

\section{Future Work}

We would like to predict the pages of a dictionary in a specific language for lookup words without any explicit manual indexing for that dictionary. We would utilize the distribution of words in the corpus of the language, distribution of the popular words of the language, and the distribution of the words in other dictionaries of that language that are already indexed. This will enable lookup in a dictionary without any manual work and will improve the accuracy with the help of crowd-sourced feedback. We would also like to explore other complex languages and their specific problems to incorporate appropriate solutions for them. Our current implementation uses a local annotation mechanism, but we would like to leverage the Open Annotation standards to expand the scope of annotations, enable open sharing, and linking across various Web services. We would like to extend the functionality of Dictionary Explorer to make it a general purpose book reader that supports indexing with the help of table of contents and annotated keywords and enables user feedback mechanism to annotate and digitize scanned books. We would also like to develop a separate client or extend the interface of our current implementation that can leverage the location indexing and annotation features for various types of archived scanned documents (such as multi-column news papers, magazines, and government records etc.) to enable lookup in them. This can be very helpful for languages that do not have good OCR support, but have active presence on the Web.

\section{Conclusions}

We have identified that general purpose online book readers are not suitable for scanned dictionaries in two ways. First, if the dictionary is in a language which is not well supported by an OCR engine used for digitizing scanned images then searching for a word yields nothing. Second, if the dictionary is well supported by an OCR engine then searching for common words returns too many results, because it cannot distinguish main words from their occurrences in the definition of other words, which is not desired for dictionary lookup. We proposed a progressive approach of indexing scanned pages of a dictionary that enables direct access to appropriate pages on lookup. Initially a sparse index is created for the dictionary that requires very little effort and makes it possible to directly jump to the page where lookup word is possibly present. In the next phase a full index is built that ensures if and where the lookup word exists in the dictionary. Finally markers are placed to precisely locate lookup words on the scanned pages. We further improve the accessibility of the dictionary by allowing annotations and fielded digitization of the dictionary words. We have implemented an application called Dictionary Explorer and utilized our technique to index various monolingual and multilingual dictionaries. We have evaluated our approach by estimating the time required for various stages of indexing and examining various trade-offs in our implementation. We have achieved a speed of over 20 pages per minute per person for sparse indexing and about 10 words per minute per person for full indexing.

\section{Acknowledgements}

We would like to thank Ayesha Aziz for taking responsibility of getting one of the biggest classical Urdu dictionaries fully indexed and providing missing scanned pages. Thanks to Shamshad and all the other UrduWeb Digital Library members who have contributed in the indexing process of various dictionaries.

\bibliographystyle{abbrv}
\bibliography{farhang}

\balancecolumns

\end{document}